# Deposit subscribe Prediction using Data Mining Techniques based Real Marketing Dataset

Safia Abbas
Computer Science Department,
Faculty of Computer and Information Sciences, Ain Shams University, Cairo, Egypt

## ABSTRACT
Recently, economic depression, which scoured all over the world, affects business organizations and banking sectors. Such economic pose causes a severe attrition for banks and customer retention becomes impossible. Accordingly, marketing managers are in need to increase marketing campaigns, whereas organizations evade both expenses and business expansion. In order to solve such riddle, data mining techniques is used as an uttermost factor in data analysis, data summarizations, hidden pattern discovery, and data interpretation. In this paper, rough set theory and decision tree mining techniques have been implemented, using a real marketing data obtained from Portuguese marketing campaign related to bank deposit subscription [Moro et al., 2011]. The paper aims to improve the efficiency of the marketing campaigns and helping the decision makers by reducing the number of features, that describes the dataset and spotting on the most significant ones, and predict the deposit customer retention criteria based on potential predictive rules.

## 1. INTRODUCTION
Business intelligence is a recent term that concerns with using the information space and intelligent mechanisms to support business managers' decisions [1,2]. Since business organizations including banking sectors yields tones of records and transactions every day, the most suitable intelligent mechanisms that can handle such vast growth of data set and information is data mining techniques (DM)[ 3, 4].

Data mining is known as the process of monitoring new and innovative information from vast amount of data sets by discovering hidden and unknown relationships between features that are entailed in the data records, spotting on the interesting events and buried patterns, summarizing the information space to extract predictive decision rules, discriminating the information space into set of objects and minimizing the features the describes the information space [5, 6].Accordingly, DM can be used to aid decision makers in banking sector to confront the economical pretense by avoiding risky transactions that cause bank attrition and increasing the customer retention incentives to raise the bank revenues [7, 8].

So, this paper focuses on two classification DM techniques, decision tree (DT) and rough set theory (RST) using real world data set collected from Portuguese marketing campaign and concerned about customer deposit subscription [9]. The paper utilizes those two techniques aiming to (i) find the redact set; the minimal set of attributes that can discriminate between objects with respect to the approximations of the information space, (ii) extract predictive rules with accuracy to aid in decision making and avoiding risks, (iii) compare between the early analysis techniques and DM techniques results.

The rest of the paper is organized as follows: section 2 introduces the data set description and its implantation, section 3 explains the preliminaries of both RST and DT as mining techniques, section 4 exposes the implementation and the obtained computational results, section 5 discusses and analyzes the differences between the ordinary analysis methods and the DM methodologies, finally section 6 shows the conclusion.

## 2. DATA SET TERMINOLOGY
In this research, we use a real dataset which was collected from a Portuguese bank that used its own contact-center to do direct marketing campaigns in order to motivate and attract the deposit clients. The dataset is related to 17 marketing campaigns and corresponding to 79354 contacts. The telephone and the internet were the central marketing channel, in which, an attractive long-term deposit application, with good interest rates, was offered [10, 11, and 12].

There are two datasets:

1- Bank-full.csv that contains various examples corresponding to 45211 objects and ordered by date.

2- bank.csv that holds 10% of the examples (4521 record), randomly selected from bank-full.csv. However, it contains almost the all possible varieties for the attributes' values and objects instances.

The bank.csv data set was firstly used in the implementation phase as a test database; however it has been implemented in the form of relation database as seen below in the database implementation subsection

### 2. 1 Data Set Description
The dataset consists of one table with 16 non- empty conditional attributes and one decision attribute, where:

- Age    : the age of the customer
- Job    : type of job (categorical)
- marital : marital status (categorical)
- education: the education level (categorical)
- Default: has credit in default?
- Balance : average yearly balance
- Housing: has housing loan?

- Loan : has personal loan?
- Contact : last contact of the current campaign (categorical)
- day : last contact day
- month : last contact month
- duration : last contact duration in seconds
- campaign : number of contacts performed during this campaign for this client includes last contact.
- pdays : number of days that passed by after the client was last contacted from a previous campaign
- previous : number of contacts performed before this campaign for this client
- poutcome : outcome of the previous marketing campaign (Categorical)

**Output attribute (desired target):**

- Deposit_s: has the client subscribed a term deposit?

The attributes types various among continues, categorical, binary and discrete. Where, categorical type means that its value is limited between several choices as described in Table1.

**Table 1: The data information**

| | Data information | | |
|---|---|---|---|
| No. | name | Type | Categorical values |
| 1 | Age | Discrete | |
| 2 | Job | Continues | admin,unknown,unemployed,management,housemaid,entrepreneur,student,blue collar,self-employed,retired,technician,services |
| 3 | Marital status | Continues | Married,Divorced (means divorced or widowed),Single |
| 4 | Education | Continues | Unknown,secondary,primary,tertiary |
| 5 | Default | Binary(Y/N) | |
| 6 | Balance | Discrete | |
| 7 | Housing | Binary(Y/N) | |
| 8 | Loan | Binary(Y/N) | |
| 9 | Contact_Type | Continues | Unknown,telephone,cellular |
| 10 | Day | Discrete | |
| 11 | Month | Discrete | |
| 12 | Duration | Discrete | |
| 13 | Campaign | Discrete | |
| 14 | Pdays | Discrete (-1 means client was not previously contacted) | |
| 15 | Previous | Discrete | |
| 16 | Poutcome | Continues | unknown,other,failure,success |
| 17 | Deposit_s | **Binary(Y/N)** | |

## 2.2 Database Implementation

Since it is very difficult to use the raw data set as one text file in the mining phase, the given dataset has been implemented in the form of relational database (RDB), in which, the contained information are divided into three types, and each type holds its information in a separate information table as follows:

- Client_info: contains all clients' information including the decision attributes and it has a sub table "Client_info" that holds the categorical values.

    o Job_Category: sub table contains all categorical values for job attribute

- Bank_info : contains all bank information related to each client.

- Campaign_info: contains the information of the 17 campaigns for each client

After then, tables are created and linked through many relations as shown in figure 1. Accordingly some attributes have been added and others have been changed in type. As example, the "Job" feature, as shown in fig.1, has been changed from type "continues\categorical" into "Discrete", and considered as a foreign key to table "Job_Category", in which each job is encoded by fixed and unique index. Despite there are 6 attributes contain a "categorical" values, only the "job" attribute treated as a foreign key that holds a sub table for the categorical values, this done due to the big variance of the entailed values comparing with the other categorical attributes.

Consequently, the number of attributes in the RDB is increased and some attributes' type are changed. However, the set of conditional attributes and the decision attribute remains the same during the mining phase. Figure 1, clarifies the tables and their relations.

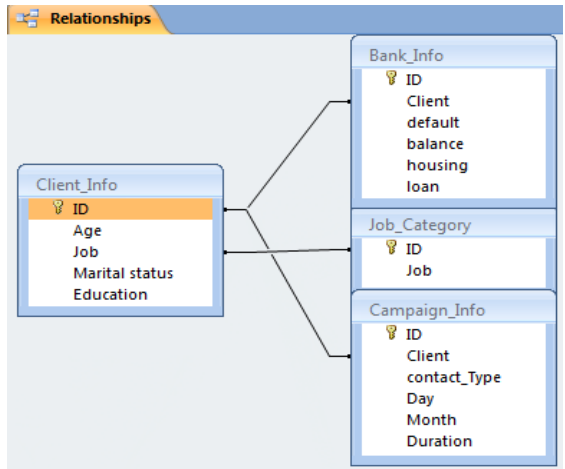

**Figure 1: Tables and relations**

## 3. DATA MINING TECHNIQUES
This section explains the main concepts and the preliminaries of both the rough set theory and the decision tree. Both techniques are discussed in details in order to be further used as mining techniques based the marketing dataset.

### 3.1 Rough Set Theory (RST) Preliminaries
This subsection recalls some necessary definitions from RST that are used in reduction process, for more details and formal definitions about the rough set theory see [16, 17, and 18].

In rough set theory, sample objects of interest are usually represented by a table called an information table. Rows of an information table correspond to objects and columns correspond to objects' features.

**Definition 1** (information system) an information system is an ordered pair

$S = <U, C \cup D>$ and $C \cap D = \emptyset$, where $U$ is a non-empty, finite set called the universe set of objects or instances, C is a non-empty finite set of conditional attributes or features, D is a non-empty finite set of decision attributes.

**Definition 2** (Indiscernibility relation) let $S = <U, C \cup D>$ be an information system, $\forall R \subset C \ \exists$ an equaivalence relation $IND(R)$ called an **indiscernibility relation**, where $IND(R)=\{(x,y) \in U \times U: a(x)=a(y)$ for every $a \in R\}$.

**Definition 3** (Approximations) Given an information system $S = <U, C \cup D>$, let $X \subseteq U$ be a set of objects and $R \subseteq C$ a selected set of attributes. The **lower approximation** of X with respect to R is $\underline{R}(X)=\{u \in U:[u]_R \subseteq X\}$. The **upper approximation** of X with respect to R is $\overline{R}(X) = \{u \in U: [u]_R \cap X \neq \emptyset\}$, If X is **R-definable** then $\underline{R}(X) = \overline{R}(X)$ otherwise X is R-Rough.

**Definition 4** (discernibility matrix) The **discernibility matrix** of an information system $S = <U, C \cup D>$ is a symmetric $n \times n$ where $|U| = n$ matrix with entries $m_{ij}$ entries. Such that $m_{ij}= \{a \in C : a(m_i) \neq a(m_j)\}$ and $m_{ij}=\emptyset$ otherwise.

**Definition 5** (CORE) Let M be a discernibility matrix of an information system $=<U, C \cup D>$, then the set R is a **CORE** if and only if $R \subset C \land \exists \gamma \in M, \gamma = R$.

### 3.2 Decision tree (DT) Preliminaries
Decision tree is a common and popular data mining technique that uses the tree hierarchy for data classification and rule inductions. The internal nodes of the tree represent the attributes' tests, the branches hold the resulted tests' values, and the leaf nodes represent the class labels for the decision attributes. For new object classification, a path for the attribute values of that object are examined according to the decision tree nodes and branches, starting from the root node till reach to the leaf node that holds the class label, such class label is considered as the class prediction for the new object [13,14, and 15].

Despite the decision tree classification is widely used, and many decision algorithms are developed in order to obtain the minimal set of attributes that are needed to build a more certain decision tree, the problem of finding the best tree representation for specific information space is still NP-hard problem [19, 20]. The tree building is mainly depends on the divide-and-conquer algorithm [21] and all current decision tree building algorithms are heuristic; the heuristic is to select and split the attribute with maximum gain ratio based on the associated information.

The gain ratio of the splitting attribute A is calculated based on the information theory as follows:

$$GAIN\_Ratio_{(A)} = \frac{H(DS) - H(DS|A_i)}{H(A_i)} \quad (1)$$

Where DS is the information space (Dataset), and A is the attribute under selection,

$$H(DS) = -\sum_{i=1}^{|D|} p(d_i) \log_2 p(d_i) \quad (2)$$

Where |D| is the number of classes partitioned by the decision class of attributes, di is the different decision values, p(di) is the probability of each label class value.

$$H(DS|A_i) = \sum_{i=1}^{v} \frac{|C_i|}{|DS|} H(C_i) \quad (3)$$

Where Ci is the class that partitioned by the value i of the selected attribute A,

$$H(A_i) = -\sum_{i=1}^{v} \frac{|D_i|}{|DS|} \log_2 \left(\frac{|D_i|}{|DS|}\right) \quad (2)$$

Where H(Ai) represents the information generated by splitting the information space DS into v partitions, corresponding to the v tests on attribute A.

## 4. IMPLEMENTATION AND COMPUTATIONAL RESULTS
This section represents the implementation of both RST and DT as data mining techniques using the previously described RDB in order to get the most significant features that describe the information tables, and extract the potential effective prediction rules.

## 4.1 RST Reduction and Decision Rules

Despite rough set theory suffers from different issues, such as the complexity time and memory space that are needed to build and perform Pawlak discernibility matrix [18], RST still the most efficient method that decides whether some of the attributes are redundant with respect to the decision class. The discernibility matrix is exploited in finding the real reduct or minimal subset of attributes, which in turn, preserves the classification (lower and upper) power for the original dataset/information table.

In this subsection, rough set concepts have been implemented on the test dataset (bank.csv) using the RDB information tables. The indiscernibility relation was implemented using the 16 conditional set of features, and the results are used in building Pawlak's discernibility matrix [18,19] as shown in figure 2.

**Figure 2: Discernibility matrix**

By Pawlak [17, 18], a reduct must satisfy an independent condition, that is, a set R is independent if $\forall r \in R \neq \emptyset, IND(R) \neq IND(R - \{r\})$ where $R \subset C$, C is the set of conditional attributes (features), using the information table without the decision attributes, and IND is the indiscernibility relation.

Accordingly, Pawlak's matrix, as shown in section 3.1, has been built by comparing objects' features for each object in the universe set U (set of all objects), and registers the differences. After then, the effect of each feature (r) on the classification power based the originalinformation table istested solely, if it satisfies the previously independentrelation, that affects the approximation classes, it is considered as mandatory ones and added to the minimal reduction set (R).

Surprisingly, the matrix reflects that the Age, Balance and Duration features are the most significant independent attributes that can be used to describe the data, and preserve the approximations, instead of the mention 16 features that are used in the original description of the information dataset. Moreover, the reduct set R={Age, Balance, Duration} can be considered as the CORE.

Accordingly, huge number of classification rules can be inferred based on the CORE set. However, we focus on the rules, in which, the error percentage is less than 25% for the test data, then we measure the accuracy of these decision rules based on the real dataset (bank-full.CSV data records). The following are some significant classification rules associated with the accuracy rate that are obtained using the test dataset and verified using the original dataset:

- If $(18 \leq age \wedge age < 25) \wedge (balace \leq 0)$ then $Deposit\_s = No$

    *for accuracy* 89.1%

- If$(18 \leq age \wedge age < 25) \wedge (balace \leq 0) \wedge (Duration \leq 95\ seconds)$ then $Deposit\_s = No$

    *for accuracy* 96.2%

- If$(25 \leq age \wedge age < 30) \wedge (balance \leq 0) \wedge (Duration \leq 95\ seconds)$ then $Deposit\_s = No$

    *for accuracy* 99.2%

- If$(30 \leq age \wedge age \leq 60) \wedge (Duration \leq 95\ seconds) \wedge (Balance \leq 0)$ then $Deposit\_s = No$ *for accuracy 99.9%*

- If$(30 \leq age \wedge age \leq 60) \wedge (Duration > 95\ seconds) \wedge (Balance \leq 0)$ then $Deposit\_s = No$ *for accuracy 92.9%*

- if$(age \leq 60) \wedge (300 < Duration < 600)$ then $Deposit\_s = No$

*for accuracy 82.48%*

- if$(Duration \leq 211)$ then Deposit_s=No *for accuracy 97.13%*

- if$(Duration \leq 645) \wedge (age \leq 60) \wedge (Balance \leq 2469)$ then $Deposit\_s = No$ *for accuracy 92.78%*

## 4.2 DT Induction and Classification

Decision tree is a widely used data mining technique that depends on converting the information sample space into a predictive tree structure that can be used to classify unknown cases using the pruning paths. Many algorithms have been devoted to achieve the best tree structure [19, 20] depending on the greedy mechanism (maximum gain ration) for selecting and splitting nodes. Regarding the provided marketing dataset, the GAIN_Ration for each attribute has been calculated and arranged in a descending order as shown in table 2.

**Table2: Attributes Gain_Ratio in a descending order**

| Attribute | GAIN_Ratio | Attribute | GAIN_Ratio |
|---|---|---|---|
| **Duration** | 0.10811967 | Housing | 0.00782731 |
| **Poutcome** | 0.03758116 | Balance | 0.00533738 |

| pdays | 0.03553361 | Loan | 0.0041129 |
|---|---|---|---|
| month | 0.0299014 | Campaign | 0.00304631 |
| Contact | 0.0299014 | Martial | 0.00297254 |
| Previous | 0.01622639 | Education | 0.00236554 |
| Job | 0.00999086 | Default | 0.00000121 |
| Age | 0.00971603 | Day | Almost zero |

Accordingly, the attribute with the maximum gain will be the root node, in other words, Duration will be the root node of the tree. After then, based the root test values, and C4.5 classifier [20l], the obtained decision tree has 104 leaves with size 146. The first level on the tree can be split into two branches (Duration >211 seconds and Duration <=211 seconds) as shown in figure 3.

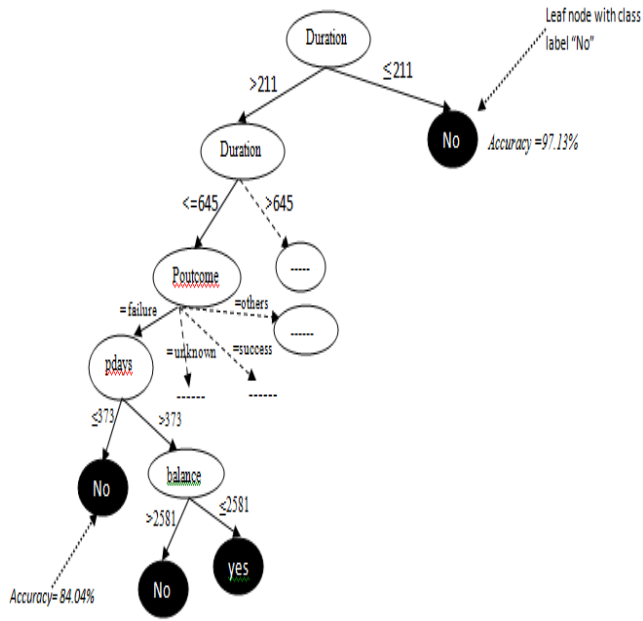

**Figure 3: the decision tree using classifier C4.5**

Decision tree enjoys several advantages, such that, it is easy to be built, easy to understood, easy to be pruned for predicting new cases, and can be converted into a set of decision rules. As a result for the tree representation in figure 3, many decision rules can be extracted as follows:-

- **If** (Duration ≤ 211) **then**Deposit_s = NO   **For accuracy = 97.13%.**

- **If** (211<Duration <=645) ∧ (Poutcome="failure") ∧ (Pdays <=373) **then** Deposit_s= "No"
  **For accuracy = 84.04%**

- **If** (Duration >645) ∧ (martial="single" ) **then** Deposit_s= "Yes"

  **For accuracy = 61.38%**

- **If** (211<Duration <=645) ∧ (Poutcome="unknown") ∧ (age <=60)∧ (contact ="cellular" )∧ (month="may") **then** Deposit_s= "No"

**For accuracy=86.27%**

- **If** (211<Duration <=645) ∧ (Poutcome="unknown") ∧ (age <=60)∧ (contact ="cellular" )∧ (month="aug") **then** Deposit_s= "No"

**For accuracy=87.58%**

- **If** (211<Duration <=645) ∧ (Poutcome="unknown") ∧ (age <=60)∧ (contact ="unknown" ) **then** Deposit_s= "No"

**For accuracy=96.55%**

- **If** (211<Duration <=645) ∧ (Poutcome="unknown") ∧ (age <=60)∧ (contact ="telephone" ) **then** Deposit_s= "No"

**For accuracy=80.00%**

## 5. ANALYSIS AND DISCUSSION

Data analysis is a very important process that is needed to facilitate data interpretations and aided indirectly the decision makers by providing need knowledge for analysis [16]. Graphics and visualization are used in data analysis to facilitate data understanding.  As example, in figure 4, Weka version 3-6-11 has been used to visualize the provided market campaign data , using bar chart, where the blue color related to the class of customers who did not subscribe for a deposit account in the bank after the campaign ended (Deposit_s = "no"), while the red color indicates the customers who subscribe for a deposit account (Deposit_s = "yes")

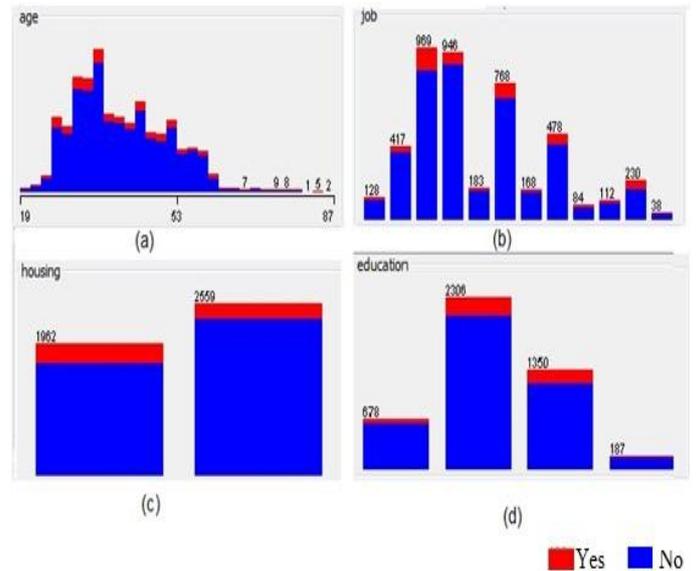

**Figure 4: Sample Space Visualization for Customers Dataset**

Despite data analysis using charts, curves, and plots seems understandable, it is very difficult to aid in predicting the needed decision with error ratios in order to minimize the risk ration. This kind of data representation depends mainly on the decision maker skills and experiences to admit a specific decision without a certainty or accuracy to support or refute such decision. So, various data mining approaches are now used for reasoning about knowledge base and provide ratios and indicators that aids in decision making

In this paper, rough set theory and decision tree have been implemented as data mining techniques based real banking data aiming to minimizing management risk, raise customer retention ratios, provide predictive and decision rules to approve or disapprove decisions based on the previous existing knowledge.

RST implementation showed that the most significant set of attributes that can describe the information space is {Balance, Age, Duration}, which is known as the Reduct set. Moreover, such set considered as a CORE set, in which the data can be analyzed using the three attributes together to differentiate between objects and preserve the approximations in the information space. Accordingly, if the analyses focused on the attributes entailed in the Reduct set, significant decision and predictive rules are extracted easily. As example, figure 5, shows the information space visualization based on the duration attribute.

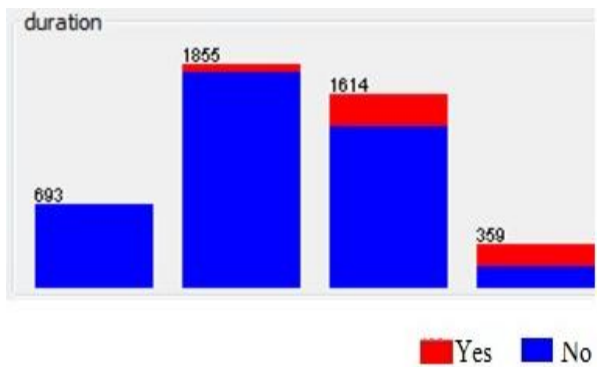

**Figure 5: Sample Space visualization based Duration feature**

As shown, the information space is divided into four classes,

- **Class1:** "Duration <= 75.5 seconds", the number of objects support these values are 693 objects, only one object belongs to the decision class "yes"; "Deposit_s="yes".

- **Class 2:** "75.5<Duration<=211.5 seconds", the number of objects support these values are 1855, only 72 objects belong to the decision class "yes".

- **Class 3:** "211.5 < Duration <=645.5 seconds", the number of objects support these values are 1614, only 268 objects belong to the decision class "yes".

- **Class 4:** "Duration>645.5 seconds", the number of objects supports these values are 359, with 180 objects belong to the decision class "yes".

Thus, the following decision rules can be extracted easily:

- If (Duration<=75.5) then Deposit_s="No" for accuracy **99.85%**.

- If (75.5< Duration <=211.5) then Deposit_s="No" for accuracy **96.11%**.

- If (211.5< Duration <=645.5) then Deposit_s="No" for accuracy **83.39%**.

- If (Duration>645.5) then Deposit_s="Yes" for accuracy **50.14%**.

More rules can be deduced and extracted from the information space based the CORE set, as example of such rules see subsection 4.1.

Finally, DT mechanism showed that some attributes have more gain values than the others, accordingly, if the data analysis focused on the maximum gain ration features, significant rules can be extracted. For example, as seen in subsection 4.2, the attribute "Poutcome" is the second feature that has the maximum gain ration="0.03758116". So, the information space has been analyzed based the categorical attribute "Poutcome", and the analysis visualization has been shown in figure 6. As shown, the information space divided into four classes based the categorical values of the selected feature.

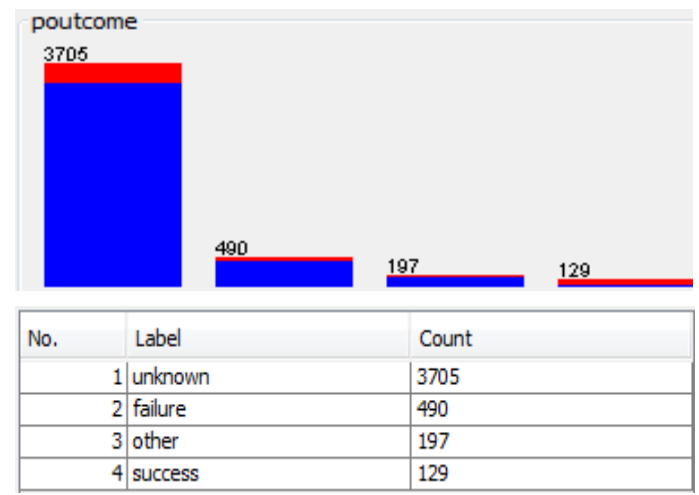

**Figure 6: Sample Space visualization based Poutcome feature**

- **Class 1:** the (poutcome="unknown"), the number of objects support these values are 3705 objects, only 337 objects belongs to the decision class "yes".

- **Class 2:** the (poutcome="failure"), the number of objects support these values are 490 objects, only 63 objects belongs to the decision class "yes".

- **Class 3:** the (poutcome="other"), the number of objects support these values are 197 objects, only 38 objects belongs to the decision class "yes".

- **Class 4:** the (poutcome="success"), the number of objects support these values are 129 objects, with 83 objects belongs to the decision class "yes".

**Accordingly, the following decision or predictive rules can be extracted easily:**

- **If** (poutcome="unknown"**) then Deposit_s="No"** for accuracy **90.90%.**

- **If** (poutcome="failure"**)        then Deposit_s="No"** for accuracy **87.14%.**

- **If** (poutcome="other"**)        then Deposit_s="No"** for accuracy **80.71%.**

- **If** (poutcome="success") **then Deposit_s="Yes"** for accuracy **64.34%.**

More specified rules, which entailed combined features and ranges, can be deduced and extracted from the tree structure that build based on the gain ratios and using the C4.5 classifier as shown in subsection 4.2.

## 6. CONCLUSION

Reasoning about banking and marketing knowledge base is one of the most challenging issues that appears due to the vast growth of data records amount and transactions [K. Chitra 2013]. Accordingly, various data mining approaches have been presented to enable better decision making. In this paper we focus on decision tree and rough set theory as their abilities for classification of new objects. Real marketing banking data, which has been obtained from Portuguese marketing campaign related to bank deposit subscription, is used in both techniques for extracting significant decision rules, and discovering the significant features that discriminate between objects.

Rough set theory implantation showed that the set of features that describes the data set can be discriminated from 16 features into the 3 features predominant set {age, Duration, Balance}, which considered as a CORE of that dataset. Thus, huge number of more valuable decisions and predictive rules can be extracted based on such CORE set, these rules in turn, helping decision makers to have a provision about the acquiring and targeting customers, make faster and better decisions about loan approvals, and minimize the management risk by the accuracy associated with the extracted rules. So, instead of analyzing data set based 16 features, which yield an intractable process for extracting meaningful patterns, only the CORE set that entails 3 features is used.

Moreover, the Decision tree approach has been implemented on the same data set, gain ratio for each attribute has been calculated and the C4.5 classifier has been used in classification process. The gain ratios showed that the "Duration" feature has the maximum gain ratio, however, the "age" feature was the 8th in the gain, and the "balance" was the 10th. DT also provides a huge number of decision and predictive rules using the 16 features associated with accuracy. Some of the DT extracted rules are more summarized than those rules extracted by the RST and others are not. As example the first rule extracted by the DT in section 4.2 is simpler than those in section 4.1.

Therefore, despite the decision tree is easy to be implemented as a classifier, and the hardness of the rough set theory implementation, RST yields a better summarization to the data set due to the feature reduction process that achieves the best minimal set of features to describe the information space and preserve its approximations.